\documentclass[conference]{IEEEtran}
\IEEEoverridecommandlockouts
\usepackage[numbers,sort&compress]{natbib}
\usepackage{amsmath,amssymb,amsfonts}
\usepackage{algorithmic}
\usepackage{subfigure}
\usepackage{graphicx}
\usepackage{textcomp}
\usepackage{tabu}
\makeatletter
\renewcommand{\@IEEEsectpunct}{\ \,}
\makeatother
\def\BibTeX{{\rm B\kern-.05em{\sc i\kern-.025em b}\kern-.08em
    T\kern-.1667em\lower.7ex\hbox{E}\kern-.125emX}}
\begin{document}

\title{SdcNet: A Computation-Efficient CNN \\
for Object Recognition
}

\author{\IEEEauthorblockN{Yunlong Ma}
\IEEEauthorblockA{Department of Electrical and Computer Engineering\\
Concordia University\\
Montreal, Canada\\
m\_yunlon@encs.concordia.ca}
\and
\IEEEauthorblockN{Chunyan Wang}
\IEEEauthorblockA{Department of Electrical and Computer Engineering\\
Concordia University\\
Montreal, Canada\\
chunyan@ece.concordia.ca}
}

\maketitle

\begin{abstract}

Extracting features from a huge amount of data for object recognition is a challenging task. Convolution neural network can be used to meet the challenge, but it often requires a large amount of computation resources. In this paper, a computation-efficient convolutional module, named SdcBlock, is proposed and based on it, the convolution network SdcNet is introduced for object recognition tasks. In the proposed module, optimized successive depthwise convolutions supported by appropriate data management is applied in order to generate vectors containing high density and more varieties of feature information. The hyperparameters can be easily adjusted to suit varieties of tasks under different computation restrictions without significantly jeopardizing the performance. The experiments have shown that SdcNet achieved an error rate of 5.60\% in CIFAR-10 with only 55M Flops and also reduced further the error rate to 5.24\% using a moderate volume of 103M Flops. The expected computation efficiency of the SdcNet has been confirmed.
 
\end{abstract}


\section{Introduction}

Object recognition is widely used in various applications such as autopilot \cite{autopilot} and security systems \cite{deepface}. Extracting various features related to the objects from diverse backgrounds is a critical challenge. The normal procedure of object recognition contains three steps, pre-processing, feature extraction and classification. 

The feature extraction can be done by filtering-based methods, such as Wavelet \cite{wavelet} and SIFT \cite{SIFT}. SVM \cite{SVM} and Adaboost \cite{Adaboost} are often used for classification. Such processing methods are usually computation-efficient, however, they have limitations in case of a huge number of variations in object features.

To deal with the situation, machine learning approaches, in particular convolution neural network(CNN), have noticeable advantages. It uses a large number of samples to progressively determine the system parameters in order to detect various object features. The networks such as VGG \cite{VGG} and ResNet \cite{ResNet} have been reported to solve complex object recognition problems. Normaly, CNN requires a large number of layers, which, in consequence, needs a large number of parameters and huge computation volume, to achieve a good performance. Research on computation-efficient architectures including computation optimization \cite{mobilenetv2,shufflenet} and various methods of network pruning  \cite{VGG-16,VGG-19,ResNet56}, has been in progress. In MobileNetV2 \cite{mobilenetv2} and ShuffleNet \cite{shufflenet}, depthwise convolutions are used in their modules to improve the computation-efficiency. Architecture Xception \cite{Xception}, a linear stack of depthwise convolution layers with residual connections, resulted in some gains in classification performance on the ImageNet dataset.

In convolution neural networks, different modes of convolutions transform the properties of input data in different ways. It's important to control various data of different nature for appropriate modes of convolutions to extract features of different orders. Based on this idea, we propose, in this paper, a convolution module, named SdcBlock, and a CNN architecture, named SdcNet, with a view to reducing significantly the computation volume without sacrificing the recognition quality. The SdcBlock, in which successive depthwise convolutions supported by appropriate data management are applied, is specifically designed for the computation with different types of data. The block is modularized to facilitate its applications in varieties of networks. 

\section{Proposed Method}

\begin{figure*}[!ht]
\centering
\subfigure[SdcBlock(Stride=1)]{
\begin{minipage}[t]{0.33\linewidth}
\centering
\scalebox{0.72}{\includegraphics{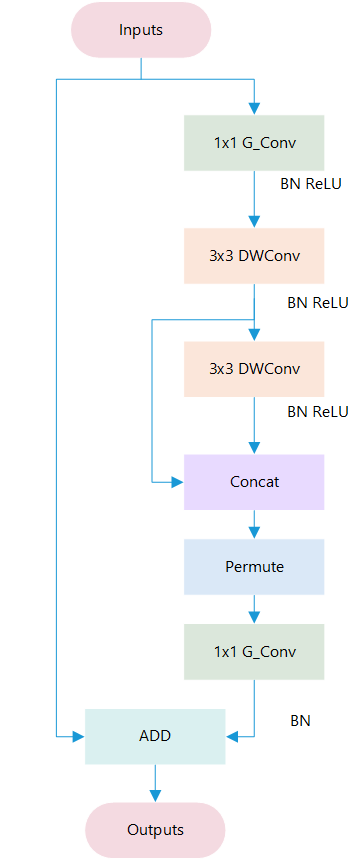}}
\label{fig:side:a}
\end{minipage}}
\subfigure[SdcBlock-S2(Stride=2)]{
\begin{minipage}[t]{0.33\linewidth}
\centering
\scalebox{0.72}{\includegraphics{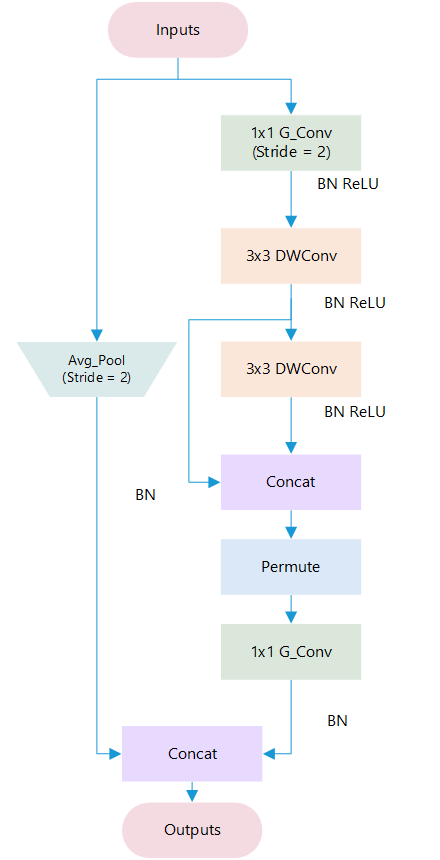}}
\label{fig:side:b}
\end{minipage}}
\subfigure[SdcBlock-S2-F(Stride=2)]{
\begin{minipage}[t]{0.25\linewidth}
\centering
\scalebox{0.72}{\includegraphics{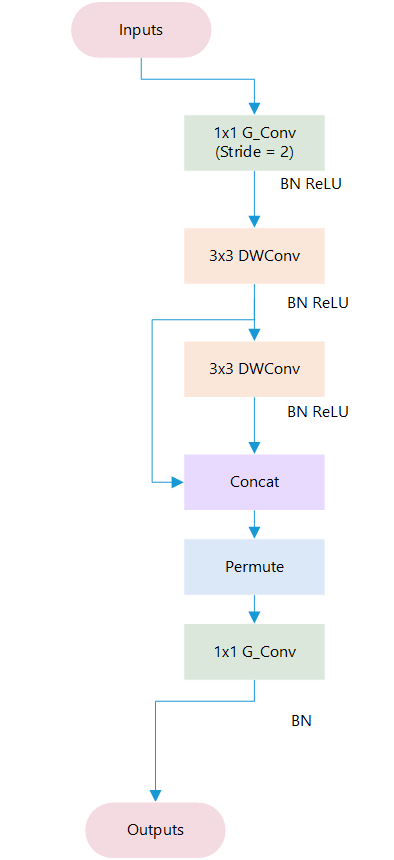}}
\label{fig:side:c}
\end{minipage}}
\caption{SdcBlock Modules. G\_Conv represents group convolution and DWConv represents dethwise convolution.}
\end{figure*}

Feature extraction by CNN is performed by means of progressive filtering through a good number of convolution layers. In each of the layers, new feature vectors are generated, based on a large volume of input data, in a way that the information relevant to the object features is extracted, composed, strengthened, and/or concentrated, while filtering out those irrelevant. Because of rich variations in the features, a large number of filtering kernels are often used in a single layer to increase the chance of extracting different features, which will certainly increase computation complexity, but not necessarily the concentration of the relevant feature information in the generated vectors. 

To build effective convolution layers with maximized capacity of extracting critical feature information, it's important to look into the different convolution modes and to direct the data to the appropriate convolution layers. In general, an input data of $N_I$ channels can be transformed to an output data of $K$ channels by a convolution with $K$ kernels. The following  modes can be the most commonly used.

\begin{itemize}
\item Standard convolution \cite{StandarConv}. In this mode, each of the $K$ convolution kernels is applied to all the $N_I$ input channels to generate one output channel.
\item Group convolution(G-Conv) \cite{Alexnet,shufflenet,ResNext}. The $N_I$ input channels is divided into $g$ groups, and $K$ convolution kernels are also divided into $g$ sets. Each group of input data is convolved with a set of $K/g$ kernels. The standard convolution can be seen as the specific case, with $g = 1$, of group convolution.
\item Depthwise convolution(DW-Conv) \cite{DepthWise}. It is, in fact, the conventional 2-D spatial convolution. In the context of CNN, it can be seen as another special case of group convolution, in which $g$  = $N_I$, \emph{i.e.}, one channel per group, and each of the $K$ kernels is applied only to one input channel, requiring $g$ = $N_I$ = $K$.
\end{itemize}

With given $N_I$ and $K$, the standard convolution is the most computation-demanding, as it generates each of the output vectors based on the data sampled from all the input channels. On the contrary, the depthwise mode is the least computation-demanding, and each of its output vectors is produced exclusively based on the data of a single channel. This exclusivity can facilitate the control of the data flow of individual channels. By using a preceding $1$x$1$ convolution to organize its input data channels, the depthwise convolution can also process data of multiple channels. 

The convolution module and architecture proposed in this paper have been designed to make the best use of the input data. It's done by means of a data management designed to optimize successive depthwise convolution results. The purpose is to generate feature vectors having higher density and more varieties of the information critical for the classification.

\subsection{Modules}

The proposed convolution module, named SdcBlock (Successive Depthwise Convolutions Block), is illustrated in Fig.1. If an input signal is composed of a large amount of data, its features can be represented by its original data form and the data resulting from filtering operations of different orders. Successive convolutions performed to the same signal can generate such feature information. Hence, the pivotal part of this module is the successive depthwise convolution to implement the principle of multiple order processing in each channel of the input data.

The module is mainly composed of three parts for the three functions, Successive Depthwise Convolution(Sdc), data preparation, and arrangement of convoluted data, respectively.

\begin{table*}[!ht]
\caption{Details of SdcNet Configurations}
\begin{tabu} to \hsize {X[l]|X[c]|X[c]|X[c]|X[c]X[c]}
\hline
Layer & Output size & Stride & Repeat Times & \multicolumn{2}{c}{{{Output channels}}} \\
 &  &  &  &   SdcNet-G4-L($g=4$) &  SdcNet-G3-S($g=3$) \\ \hline
\multicolumn{6}{c}{{{Input image data sized $32$x$32$, $3$ channels.}}} \\  \hline
G-Conv($g=3$)* & 32x32 & 1 & 1 & 36 & 36   \\  \hline
Stages 1 & 32x32  & 1 & 1 & 24  & 24\\ \hline
Stages 2 & 32x32  & 1 & 2 & 36  & 24\\ \hline
Stages 3 & 16x16  & 2 & 1 & 72 & 36\\ 
		 & 16x16  & 1 & 2 & 72  & 36\\ \hline
Stages 4 & 8x8 &  2 & 1 & 96 &  72\\ 
		 & 8x8 &  1 & 3 & 96  & 72\\ \hline
Stages 5 & 8x8 &  1 & 3 & 144 &  96\\ \hline
Stages 6 & 4x4 &  2 & 1 & 300 &  150\\ 
		 & 4x4 &  1 & 2 & 300 &  150\\ \hline
Stages 7 & 4x4 &  1 & 1 & 600 & 300\\ \hline
Avg Pool** & 1x1 & 2 & 1 & 600  & 300\\ \hline
FC*** 		 & 1x1 &    & 1 & 10 & 10 \\ \hline
Complexity**** &  & &  &        106.1M & 56.55M  \\ \hline
\multicolumn{6}{l}{{{* G-Conv stands for group convolution. The kernel size of the group convolution is $3$x$3$.}}}\\
\multicolumn{6}{l}{{{** The kernel size of the average pooling is $4$x$4$.}}}\\
\multicolumn{6}{l}{{{*** 10 is for CIFAR-10 dataset and 100 is for CIFAR-100 dataset.}}}\\
\multicolumn{6}{l}{{{**** The complexity is evaluated with FLOPs for the dataset CIFAR-10, \emph{i.e.} the number of floating-point multiplication-adds. }}}\\
\end{tabu}
\end{table*}

\begin{enumerate}
  \item \emph{Successive depthwise convolutions(Sdc).}  They are performed by two boxes, indicated by $3$x$3$ DWconv found in Fig.1 (a), used to generate the data of the first and second order filtering operations. They must be applied solely to the data of the same channel, for which only depthwise convolution is suitable. If the module is placed in the entry part of a CNN to process raw image data, the successive depthwise convolutions will generate the first and second order gradient maps in order to obtain various low-level features. If the module is placed in the middle or final parts of the convolution stages, these convolution operations will produce vectors of more dimensions and levels containing high order information. 
  
 The extracted feature information of each of the two convolutions needs to be carefully preserved. Hence, the two sets of the convolution results are concatenated, instead of being summed up.
 
 In the current version of SdcBlock illustrated in Fig.1 (a), the kernel size of the two succisive depthwsise convolutions is $3$x$3$. Batch normalization \cite{bn} and non-linear function ReLU \cite{ReLU} are applied after each of the convolutions. 

	\item \emph{Data preparation for the successive depthwise convolution.} Since the pivotal part in the proposed module is the depthwise convolutions performed in the individual input channels. It is important to prepare the input data in order to optimize the convolution results. In SdcBlock, a set of $1$x$1$ convolution kernels are applied to the input data, as illustrated in Fig.1 (a), for the preparation. By doing so, the data can be scaled to suit the succeeding convolution and, meanwhile, the number of the input data channels are expanded to match that required in the successive depthwise convolutions. If $N_I$ input channels are expanded to $E*N_I$ channels, $E$ is the expansion number, used as one of the hyperparameters.
	
	In the current version of the SdcNet, the group convolution mode is used in the first set of $1$x$1$ convolution in each block, as illustrated in Fig.1 (a), in the data preparation so that the input channels can be grouped according to their nature. Moreover, the group convolutions can reduce the computation complexity significantly, with respect to the standard convolution. It should also be mentioned that batch normalization and non-linear function ReLU are applied after the first $1$x$1$ group convolution.

	\item \emph{Data arrangement after the successive depthwise convolution.} As mentioned previously, the results from the two successive depthwise convolution operations are concatenated. The data produced by the different depthwise convolutions are placed separately in different sections of the vectors generated by the concatenation. Rearranging the vector elements so that every segment in the vectors has elements randomly taken from the two convolutions results may benefit the following operations. Thus, the results of the concatenation are permuted to mingle the results of the two depthwise convolutions. Another group convolution, the second $1$x$1$ kernel convolution illustrated in Fig.1 (a), is applied to combine the data from $2$x$E$x$N_I$ channels to $N_O$ output channels. A batch normalization is followed after the second $1$x$1$ group convolution.

\end{enumerate}

The proposed module can be varied by a choice of the hyper parameter \emph{stride}. In the basic version of the proposed SdcBlock illustrated in Fig.1 (a), it takes \emph{stride=1}. A residual operation \cite{ResNet} is applied to the inputs and the rearranged convolved results. There are three ReLUs in each SdcBlock to ensure the non-linear ability of the module and there is no ReLU after the addition. In some cases, \emph{stride=2} convolution layers can be used in order to reduce computation cost. The SdcBlock with \emph{stride = 2}, named SdcBlock-S2, is shown in Fig.1 (b). In order to compensate the eventual information lost due to the \emph{stride = 2} convolution, a  concatenation of the average pooled input data and the rearranged convolved data is performed. The final output of SdcBlock-S2 contains both the sampled input information and the successive depthwise convolution results. A variation of the SdcBlock-S2, named SdcBlock-S2-F(Feature), has also been proposed and the procedure is shown in Fig.1 (c). Compared to SdcBlock-S2, the final output data result exclusively from the successive depthwise convolutions without being combined with the input. 

\subsection{Network architecture}

A convolution neural network architecture mainly composed of a stack of SdcBlocks is named SdcNet. Two SdcNets have been designed for the CIFAR image classification and the details of the designs are presented in Table I. Each of these SdcNets is composed of seventeen SdcBlocks grouped in seven stages. The hyperparameters in each stage are made the same, except \emph{stride}.

In a SdcBlock, the hyperparameters $g$, the number of groups, and $E$, the expansion number, are used to control the computation volume. Furthermore, the volume is also closely related to the number of output channels in each stage. The two SdcNet networks, specified in Table I, differ in the number of groups and the number of output channels in stages. In SdcNet-G4-L, "G4" indicates $g = 4$ is applied to all the stages whereas $g = 3$ is used in SdcNet-G3-S. As the former has a larger number of output channels than the latter, the name of the former ends with "L", standing for "larger", and that of the latter with "S", standing for "smaller". Besides, the basic SdcBlock is used in case of \emph{stride=1} and SdcBlock-S2 is used in case of \emph{stride=2}, unless otherwise specified. 

\section{Performance Evaluation}

To evaluate the performance of SdcNet, a set of experiments have been performed with CIFAR-10 and CIFAR-100 image classification datasets.

\subsection{Experiment Conditions}

\subsubsection{Datasets.}

The CIFAR dataset \cite{Cifar} (Canadian Institute For Advanced Research) is a collection of images that are commonly used to train machine learning and computer vision algorithms. Both the CIFAR-10 and CIFAR-100 datasets contain 60000 RGB images of 32x32 pixels in 10 classes and 100 classes, respectively. For the CIFAR-10 dataset, the training set has 5000 images per class and the testing set contains 1000 randomly-selected images from each class. In the CIFAR-100 dataset, there are 500 training images per class and 100 testing images per class.

\subsubsection{Network Configurations.}

Four different versions of SdcNets have been tested. The details of the network configurations are specified in Table I and the explanation of the network names is found in Section II B. These SdcNets differ in the number of groups in each stage, the number of output channels and types of used blocks. In cases of SdcNet-G4-L-F and SdcNet-G3-S-F,  SdcBlock-S2-F are used for all the blocks in case of \emph{stride=2}. In all the four networks, $E$ is equal to 6.

\subsubsection{Training Details.}

All SdcNets are trained with stochastic gradient descent (SGD) \cite{SGD} using similar optimization hyper-test error parameters as in \cite{ResNet}. Besides, Nesterov momentum with a momentum weight of 0.9 and a weight decay of 0.0001 have been adopted. The networks are trained with mini-batch size 128 for 300 epochs. For the learning rate, a cosine shape decreasing method \cite{cosinelr} starting from 0.1 and reduces to 0.002 gradually has been used in the 300 epochs. The simple data augmentation in \cite{DataAug} has also been applied for training, four zero pixels are padded on each side, and then a $32$x$32$ crop is randomly sampled from the padded image or its horizontal flip. For testing, the original images of $32$x$32$ are used to evaluate our network.

\begin{table}
\caption{Comparison of classification error rate on CIFAR- 10 (C-10) and CIFAR-100 (C-100) with existing CNNs }
\begin{tabu} to \hsize {X[3.5,l]|X[c]|X[c]|X[c]|X[c]}
\hline
Model & FLOPs & Params & C-10 & C-100 \\ \hline
VGG-16-pruned\cite{VGG-16} & 206M & 5.40M & 6.60\% & 25.28\% \\
VGG-19-pruned\cite{VGG-19} & 195M & 2.30M & 6.20\% & - \\
VGG-19-pruned\cite{VGG-19} & 250M & 5.00M & - & 26.20\% \\ \hline
ResNet-56-pruned\cite{ResNet56} & 62M & - & 8.20\% & - \\ 
ResNet-56-pruned\cite{VGG-16} & 90M & 0.73M & 6.94\% & - \\ 
ResNet-110-pruned\cite{VGG-16} & 213M & 1.68M & 6.45\% & - \\ 
ResNet-164-B-pruned\cite{VGG-19} & 124M & 1.21M & 5.27\% & 25.28\% \\ \hline
SdcNet-G3-S-F & 56.55M & 1.09M & 5.79\% & 25.83\%  \\
SdcNet-G3-S & 55.12M & 1.04M & 5.60\% & 25.01\%  \\
SdcNet-G4-L-F & 106.1M & 2.61M & 5.27\% & 23.52\%  \\
SdcNet-G4-L & 103.3M & 2.53M & 5.24\% & 23.12\%  \\
\hline
\end{tabu}
\end{table}

\subsection{Results on CIFAR 10 and CIFAR 100}

The test result is presented in Table II. In general, the error rates achieved by SdcNets are not above 5.79\% in CIFAR-10 and 25.83\% in CIFAR-100, which are better than those given by ResNet-pruned and VGG-pruned nets under the similar computation conditions, in terms of Flops.

With the restriction of low computation cost, SdcNet can achieve an error rate of 5.60\% in CIFAR-10 with 55M Flops in comparison with 8.20\% by ResNet-56 with 62M Flops. In terms of quality of recognition, the error rate achieved by SdcNet can be as low as 5.24\% in CIFAR-10 using 103M Flops versus 5.27\% given by ResNet-164-B-pruned using 124M Flops. These results confirm that the proposed modules and networks have a better performance in terms of efficiency.

\section{Conclusion}

In this paper, a computation-efficient convolutional module, named SdcBlock, has been proposed and based on it, the convolution network SdcNet introduced for object recognition tasks. The pivotal part of the SdcBlock is the optizimized successive depthwise convolution supported by the appropriate data management to generate vectors containing high density and more variety feature information. The hyperparameters can be adjusted for varieties of tasks under different computation restrictions without significantly jeopardizing the performance. Examples of SdcNet have been designed and tested. It has been demonstrated that SdcNet achieved an error rate of 5.60\% in CIFAR-10 with only 55M Flops, and also reduced further the error rate to 5.24\% using a moderate volume of 103M Flops. The computation efficiency of the SdcNet has been confirmed although the results can be further improved by fine adjustments of hyperparameters.
   
\section*{Acknowledgment}

This research was enabled in part by support provided by Compute Canada (www.computecanada.ca). Computations were made on the supercomputers Helios and Graham managed by Calcul Québec and Compute Canada. 

\bibliographystyle{IEEEtran}
\bibliography{ref}

\end{document}